\documentclass[12pt]{article}
\makeindex
\usepackage[latin1]{inputenc}
\usepackage{amsmath, amsthm, amssymb}
\begin{document}
\newcommand{\AAA}[1]{\mathbb{A}^{#1}} 
\newcommand{\BB}[1]{\mathbb{B}^{#1}} 
\newcommand{\CC}[1]{\mathbb{C}^{#1}} 
\newcommand{\DD}[1]{\mathbb{D}^{#1}} 
\newcommand{\EE}[1]{\mathbb{E}^{#1}} 
\newcommand{\FF}[1]{\mathbb{F}^{#1}} 
\newcommand{\GG}[1]{\mathbb{G}^{#1}} 
\newcommand{\HH}[1]{\mathbb{H}^{#1}} 
\newcommand{\II}[1]{\mathbb{I}^{#1}} 
\newcommand{\JJ}[1]{\mathbb{J}^{#1}} 
\newcommand{\KK}[1]{\mathbb{K}^{#1}} 
\newcommand{\LL}[1]{\mathbb{L}^{#1}} 
\newcommand{\MM}[1]{\mathbb{M}^{#1}} 
\newcommand{\NN}[1]{\mathbb{N}^{#1}} 
\newcommand{\OO}[1]{\mathbb{O}^{#1}} 
\newcommand{\PP}[1]{\mathbb{P}^{#1}} 
\newcommand{\QQ}[1]{\mathbb{Q}^{#1}} 
\newcommand{\RR}[1]{\mathbb{R}^{#1}} 
\newcommand{\SSS}[1]{\mathbb{S}^{#1}} 
\newcommand{\TT}[1]{\mathbb{T}^{#1}} 
\newcommand{\UU}[1]{\mathbb{U}^{#1}} 
\newcommand{\VV}[1]{\mathbb{V}^{#1}} 
\newcommand{\WW}[1]{\mathbb{W}^{#1}} 
\newcommand{\XX}[1]{\mathbb{X}^{#1}} 
\newcommand{\YY}[1]{\mathbb{Y}^{#1}} 
\newcommand{\ZZ}[1]{\mathbb{Z}^{#1}} 
\newcommand{\CP}[1]{\mathbb{CP}^{#1}} 
\newcommand{\RP}[1]{\mathbb{RP}^{#1}}

\newcommand{\prodd}{\mathop{\prod}\limits}
\newcommand{\summ}{ \mathop{\sum}\limits}
\newcommand{\cupp}{\mathop{\bigcup}\limits}
\newcommand{\capp}{\mathop{\bigcap}\limits}
\newcommand{\opp}{\mathop{\bigoplus}\limits}
\newcommand{\ott}{\mathop{\bigotimes}\limits}

\newcommand{\longhook}{\lhook\joinrel\longrightarrow}
\newcommand{\hookup}{{\lower6.3pt\hbox{\cu}\kern-4.32pt\big\uparrow}}
\newcommand{\hookdown}{\raise5.3pt\hbox{\ca}\kern-4.32pt\lower3pt
  \hbox{\big\downarrow}}

\providecommand{\cf}{{\upshape cf. }}%
\providecommand{\ie}{{\slshape i.e. }}%
\providecommand{\resp}{{resp. }}%
\providecommand{\cad}{c'est-\`a-dire }%
\providecommand{\ci}{\mathcal{C}^{\infty}}%
\newcommand{\apri}{\slshape a priori }%
\newcommand{\afor}{\slshape a fortiori }%
\providecommand{\etc}{etc. }%

\newcommand{\ol}{\overline }
\newcommand{\set}{\smallsetminus}
\newcommand{\wt}{\widetilde}
\newcommand{\wh}{\widehat}
\providecommand{\ept}{\footnotesize}
\newcommand{\dis}{\displaystyle}
\providecommand{\la}{\longrightarrow}
\newcommand{\ld}{, \ldots,}
\newcommand{\simto}{\xrightarrow{\sim}}
\newcommand{\unn}{\mbox{\usefont{U}{bbold}{m}{n}1}}

\newcommand{\Ker}{\mathop{\mathrm{Ker}}\nolimits}
\newcommand{\Log}{\mathop{\mathrm {Log}}\nolimits}
\newcommand{\im}{\mathop{\mathrm {Im}}\nolimits}


\newcommand\ac{\mathcal{A}}      \newcommand\ab{\mathbb{A}}         
\newcommand\bc{\mathcal{B}}      \newcommand\bbb{\mathbb{B}}        
\newcommand\cc{\mathcal{C}}      \newcommand\cb{\mathbb{C}}         
\newcommand\dc{\mathcal{D}}      \newcommand\db{\mathbb{D}}         
\newcommand\ec{\mathcal{E}}      \newcommand\eb{\mathbb{E}}       
\newcommand\fc{\mathcal{F}}      \newcommand\fb{\mathbb{F}}         
\newcommand\gc{\mathcal{G}}      \newcommand\gb{\mathbb{G}}         
\newcommand\hc{\mathcal{H}}      \newcommand\hb{\mathbb{H}}         
\newcommand\ic{\mathcal{I}}      \newcommand\ib{\mathbb{I}}         
\newcommand\jc{\mathcal{J}}      \newcommand\jb{\mathbb{J}}        
\newcommand\kc{\mathcal{K}}      \newcommand\kb{\mathbb{K}}        
\newcommand\lc{\mathcal{L}}      \newcommand\lb{\mathbb{L}}         
\newcommand\mc{\mathcal{M}}      \newcommand\mb{\mathbb{M}}         
\newcommand\nc{\mathcal{N}}      \newcommand\nb{\mathbb{N}}         
\newcommand\oc{\mathcal{O}}      \newcommand\obb{\mathbb{O}}        
\newcommand\pcc{\mathcal{P}}     \newcommand\pb{\mathbb{P}}
\newcommand\qc{\mathcal{Q}}      \newcommand\qb{\mathbb{Q}}
\newcommand\rc{\mathcal{R}}      \newcommand\rb{\mathbb{R}}
\newcommand\scc{\mathcal{S}}     \newcommand\sbb{\mathbb{S}}
\newcommand\tc{\mathcal{T}}      \newcommand\tb{\mathbb{T}}
\newcommand\uc{\mathcal{U}}      \newcommand\ub{\mathbb{U}}
\newcommand\vc{\mathcal{V}}      \newcommand\vb{\mathbb{V}}
\newcommand\wc{\mathcal{W}}      \newcommand\wb{\mathbb{W}}
\newcommand\xc{\mathcal{X}}      \newcommand\xb{\mathbb{X}}
\newcommand\yc{\mathcal{Y}}      \newcommand\yb{\mathbb{Y}}
\newcommand\zc{\mathcal{Z}}      \newcommand\zb{\mathbb{Z}}

\newcommand\aaf{\mathbf{A}}
\newcommand\bbf{\mathbf{B}}
\newcommand\ccf{\mathbf{C}}
\newcommand\ddf{\mathbf{D}}
\newcommand\eef{\mathbf{E}}
\newcommand\fff{\mathbf{F}}
\newcommand\ggf{\mathbf{G}}
\newcommand\hhf{\mathbf{H}}
\newcommand\iif{\mathbf{I}}
\newcommand\jjf{\mathbf{J}}
\newcommand\kkf{\mathbf{K}}
\newcommand\llf{\mathbf{L}}
\newcommand\mmf{\mathbf{M}}
\newcommand\nnf{\mathbf{N}}
\newcommand\oof{\mathbf{O}}
\newcommand\ppf{\mathbf{P}}
\newcommand\qqf{\mathbf{Q}}
\newcommand\rrf{\mathbf{R}}
\newcommand\ssf{\mathbf{S}}
\newcommand\ttf{\mathbf{T}}
\newcommand\uuf{\mathbf{U}}
\newcommand\vvf{\mathbf{V}}
\newcommand\wwf{\mathbf{W}}
\newcommand\xxf{\mathbf{X}}
\newcommand\yyf{\mathbf{Y}}
\newcommand\zzf{\mathbf{Z}}

\def\bul{{\scriptstyle\scriptstyle\bullet}}
\def\cupll{\mathop{\cupp\kern-8.94pt\shortmid\kern-1pt\shortmid\kern5pt}\limits}

\newcommand{\slim}[1]{\mathop{s{-}\lim}_{#1}\limits}
\newcommand{\laa}{\mathop{\la}\limits}
\def\deff{\stackrel{{\rm def}}{=}\ }
\def\thsp{\thinspace}
\def\qq{\quad} 
\def\ld{,\ldots,}
\def\wh{\widehat}
\def\wt{\widetilde}
\def\orth{\perp}
\def\cd{\cdot}
\def\cadlag{c\`adl\`ag }
\def\la{\longrightarrow}
\def\ol{\overline}
\def\smallsetminus{{\scriptstyle \setminus}}
\def\sset{\smallsetminus}
\newcommand{\gnb}{\gamma _{_{\mbox{$\scriptstyle\nb$}}}}
\newcommand{\ve}{\mathop{\bigvee}\limits}
\newcommand{\we}{\mathop{\bigwedge}\limits}
\newcommand{\esssup}{\mathop{\rm ess\ sup}\limits}
\def\olgt{\ol{\Gamma (T)}}
\def\dis{\displaystyle}
\def\pa{\partial}
\def\indic{{\mathop{\rm 1\mkern-4mu l}}}
\let\unn=\indic 

\def\Ree{~{\Re e}}
\def\ind{\mathop{{\rm index}}\nolimits}
\def\resp{{\it resp.\/}}
 \def\Hom{\mathop{{\rm Hom}}\nolimits}
\def\loc{\mathop{{\rm loc}}\nolimits}
\def\ad{\mathop{{\rm ad}}\nolimits}
 \def\sym{\mathop{{\rm sym}}\nolimits}
\def\Dom{\mathop{{\rm Dom}}\nolimits}
\def\Ran{\mathop{{\rm Ran}}\nolimits}
\def\Tr{\mathop{\rm Tr}\nolimits}
\def\tr{\mathop{\rm Tr}\nolimits}
\def\im{\mathop{\rm Im}\nolimits}
\def\per{\mathop{\rm per}\nolimits}
\def\ker{\mathop{\rm Ker}\nolimits}
\def\det{\mathop{\rm det}\nolimits}
\def\Var{\mathop{\rm Var}\nolimits}
\def\as{{\rm a.s.\thsp}}
\def\aa{a.a.\thsp}

\def\lan{\langle}
\def\ran{\rangle}

\def\ofp{(\Omega,\fc,P)}
\def\Rp{{\rb^{+}\!}}
\def\Rn{{\rb^{n}\!}}
\def\RN{{\rb^{}\!}}
\def\lrp{L^2(\Rp)}
\def\compp#1{\stackrel{\ c}{#1}}
\def\comp#1{\stackrel{c}{#1}{ }}
\def\ps#1#2{\langle #1\, ,\, #2\rangle}
\def\croo#1#2{{\langle#1,#2\rangle}}
\def\crod#1#2{{[#1,#2]}}
\def\IS{{\scriptstyle{\mathord{\Sigma\!\!\!\!\!\int}}}}
\def\[{{\mathord{[\! [}}}
\def\]{{\mathord{]\! ]}}}
\def\ppre#1{\,{{}^{{}^{\scriptstyle p}}}\!#1}
\def\popt#1{\,{{}^{{}^{\scriptstyle o}}}\!#1}
\def\espcond#1#2{\eb\left[#1\;| \;#2\right]}
\def\esp#1{\eb\left[#1\right]}
\def\norme#1{\| #1\|}
\def\normca#1{{\| #1\|}^2}
\def\ab#1{\vert #1\vert}
\def\bab#1{\vert #1\vert}
\def\Bab#1{\vert #1\vert}
\def\seq#1{\left(#1_n\right)}
\def\vect#1{\vec{#1}}
\def\pro#1{#1_{\cdot}}
\def\tirr#1{{\left(#1\right)}_{t\in \Rp}}
\def\tir#1{#1_{\cdot}}
\def\sir#1{#1_{\cdot}} %
\def\sirr#1{{\left(#1\right)}_{s\in \Rp}}
\def\L{\Lambda}
\def\O{\Omega}
\def\F{\Phi}
\def\Y{\psi}
\def\D{\Delta}
\def\a{\alpha}
\def\b{\beta}
\def\s{\sigma}
\def\m{\mu}
\def\r{\rho}
\def\n{\nu}
\def\N{\nabla}
\def\e{\varepsilon}
\def\f{\phi}
\def\l{\lambda}
\def\g{\gamma}

\let\rF=\fc
\let\rB=\bc
\let\rO=\oc
\let\rN=\nc
\let\rG=\gc
\let\rP=\pcc
\let\rM=\mc
\let\rH=\hc
\let\rJ=\jc
\let\rI=\ic
\let\rS=\scc
\let\rC=\cc
\let\rD=\dc
\let\rE=\ec
\let\rF=\fc
\let\rK=\kc
\let\rL=\lc
\let\rN=\nc
\let\rQ=\qc
\let\rR=\rc
\let\rT=\tc
\let\rA=\ac
\let\rU=\uc
\let\rV=\vc
\let\rW=\wc
\newcommand{\ccc}{\mathcal {C}}

\let\CC=\cb
\let\EE=\eb
\let\FF=\fb
\let\HH=\hb
\let\NN=\nb
\let\RR=\rb
\let\PP=\pb
\let\QQ=\qb
\let\ZZ=\zb

\def\cstar{$C^*$-algebra}

\newcommand{\va}{\rm variable al\'eatoire}
\newcommand{\vas}{\rm variables al\'eatoires}
\renewcommand{\Im}{\mathop{\rm Im }}
\renewcommand{\Re}{\mathop{\rm Re }}
\newcommand{\ra}{\mathop{\rightarrow }}
\newcommand{\supp}{\mathop{\rm supp}}
\newcommand{\sgn}{\mathop{\rm sgn }}
\newcommand{\card}{\mathop{\rm card }}
\newcommand{\KM}{\mbox{\rm KM}}
\newcommand{\diam}{\mathop{\rm diam}}
\newcommand{\diag}{\mathop{\rm diag}}
\newcommand{\dd}{\mathop{\rm d}}
\newcommand{\id}{\mbox{\rm1\hspace{-.2ex}\rule{.1ex}{1.44ex}}
   \hspace{-.82ex}\rule[-.01ex]{1.07ex}{.1ex}\hspace{.2ex}}
\renewcommand{\P}{\mathop{\rm Prob}}
\newcommand{\V}{\mathop{\rm Var}}
\newcommand{\cps}{{\stackrel{{\rm p.s.}}{\rightarrow}}}
\newcommand{\limm}{\mathop{\rm l.i.m.}}
\newcommand{\cloi}{{\stackrel{{\rm loi}}{\rightarrow}}}
\newcommand{\bra}{\langle\,}
\newcommand{\ket}{\,\rangle}
\newcommand{\obl}{/\!/}
\newcommand{\mapdown}[1]{\vbox{\vskip 4.25pt\hbox{\bigg\downarrow
  \rlap{$\vcenter{\hbox{$#1$}}$}}\vskip 1pt}}
\newcommand{\tab}{&\!\!\!}
\newcommand{\tabb}{&\!\!\!\!\!}
\renewcommand{\d}{\displaystyle}
\newcommand{\epreuve}{\hspace{\fill}$\bigtriangleup$}
\newcommand{\demo}{{\par\noindent{\em D\'emonstration~:~}}}
\newcommand{\solu}{{\par\noindent{\em Solution~:~}}}
\newcommand{\NB}{{\par\noindent{\bf Remarque~:~}}}
\newcommand{\const}{{\rm const}}

\newcommand{\cA}{{\cal A}}
\newcommand{\cB}{{\cal B}}
\newcommand{\cC}{{\cal C}}
\newcommand{\cD}{{\cal D}}
\newcommand{\cE}{{\cal E}}
\newcommand{\cF}{{\cal F}}
\newcommand{\cG}{{\cal G}}
\newcommand{\cH}{{\cal H}}
\newcommand{\cI}{{\cal I}}
\newcommand{\cJ}{{\cal J}}
\newcommand{\cK}{{\cal K}}
\newcommand{\cL}{{\cal L}}
\newcommand{\cM}{{\cal M}}
\newcommand{\cN}{{\cal N}}
\newcommand{\cO}{{\cal O}}
\newcommand{\cP}{{\cal P}}
\newcommand{\cQ}{{\cal Q}}
\newcommand{\cR}{{\cal R}}
\newcommand{\cS}{{\cal S}}
\newcommand{\cT}{{\cal T}}
\newcommand{\cU}{{\cal U}}
\newcommand{\cV}{{\cal V}}
\newcommand{\cW}{{\cal W}}
\newcommand{\cX}{{\cal X}}
\newcommand{\cY}{{\cal Y}}
\newcommand{\cZ}{{\cal Z}}
\newcommand{\bA}{{\bf A}}
\newcommand{\bB}{{\bf B}}
\newcommand{\bC}{{\bf C}}
\newcommand{\bD}{{\bf D}}
\newcommand{\bE}{{\bf E}}
\newcommand{\bF}{{\bf F}}
\newcommand{\bG}{{\bf G}}
\newcommand{\bH}{{\bf H}}
\newcommand{\bI}{{\bf I}}
\newcommand{\bJ}{{\bf J}}
\newcommand{\bK}{{\bf K}}
\newcommand{\bL}{{\bf L}}
\newcommand{\bM}{{\bf M}}
\newcommand{\bN}{{\bf N}}
\newcommand{\bP}{{\bf P}}
\newcommand{\bQ}{{\bf Q}}
\newcommand{\bR}{{\bf R}}
\newcommand{\bS}{{\bf S}}
\newcommand{\bT}{{\bf T}}
\newcommand{\bU}{{\bf U}}
\newcommand{\bV}{{\bf V}}
\newcommand{\bW}{{\bf W}}
\newcommand{\bX}{{\bf X}}
\newcommand{\bY}{{\bf Y}}
\newcommand{\bZ}{{\bf Z}}
\newcommand{\bu}{{\bf u}}
\newcommand{\bv}{{\bf v}}
\newfont{\msbm}{msbm10 scaled\magstep1}
\newfont{\msbms}{msbm7 scaled\magstep1} 
\newcommand{\bbA}{\mbox{$\mbox{\msbm A}$}}
\newcommand{\bbB}{\mbox{$\mbox{\msbm B}$}}
\newcommand{\bbC}{\mbox{$\mbox{\msbm C}$}}
\newcommand{\bbD}{\mbox{$\mbox{\msbm D}$}}
\newcommand{\bbE}{\mbox{$\mbox{\msbm E}$}}
\newcommand{\bbF}{\mbox{$\mbox{\msbm F}$}}
\newcommand{\bbG}{\mbox{$\mbox{\msbm G}$}}
\newcommand{\bbH}{\mbox{$\mbox{\msbm H}$}}
\newcommand{\bbI}{\mbox{$\mbox{\msbm I}$}}
\newcommand{\bbJ}{\mbox{$\mbox{\msbm J}$}}
\newcommand{\bbK}{\mbox{$\mbox{\msbm K}$}}
\newcommand{\bbL}{\mbox{$\mbox{\msbm L}$}}
\newcommand{\bbM}{\mbox{$\mbox{\msbm M}$}}
\newcommand{\bbN}{\mbox{$\mbox{\msbm N}$}}
\newcommand{\bbO}{\mbox{$\mbox{\msbm O}$}}
\newcommand{\bbP}{\mbox{$\mbox{\msbm P}$}}
\newcommand{\bbQ}{\mbox{$\mbox{\msbm Q}$}}
\newcommand{\bbR}{\mbox{$\mbox{\msbm R}$}}
\newcommand{\bbS}{\mbox{$\mbox{\msbm S}$}}
\newcommand{\bbT}{\mbox{$\mbox{\msbm T}$}}
\newcommand{\bbU}{\mbox{$\mbox{\msbm U}$}}
\newcommand{\bbV}{\mbox{$\mbox{\msbm V}$}}
\newcommand{\bbW}{\mbox{$\mbox{\msbm W}$}}
\newcommand{\bbX}{\mbox{$\mbox{\msbm X}$}}
\newcommand{\bbY}{\mbox{$\mbox{\msbm Y}$}}
\newcommand{\bbZ}{\mbox{$\mbox{\msbm Z}$}}
\newcommand{\bbsA}{\mbox{$\mbox{\msbms A}$}}
\newcommand{\bbsB}{\mbox{$\mbox{\msbms B}$}}
\newcommand{\bbsC}{\mbox{$\mbox{\msbms C}$}}
\newcommand{\bbsD}{\mbox{$\mbox{\msbms D}$}}
\newcommand{\bbsE}{\mbox{$\mbox{\msbms E}$}}
\newcommand{\bbsF}{\mbox{$\mbox{\msbms F}$}}
\newcommand{\bbsG}{\mbox{$\mbox{\msbms G}$}}
\newcommand{\bbsH}{\mbox{$\mbox{\msbms H}$}}
\newcommand{\bbsI}{\mbox{$\mbox{\msbms I}$}}
\newcommand{\bbsJ}{\mbox{$\mbox{\msbms J}$}}
\newcommand{\bbsK}{\mbox{$\mbox{\msbms K}$}}
\newcommand{\bbsL}{\mbox{$\mbox{\msbms L}$}}
\newcommand{\bbsM}{\mbox{$\mbox{\msbms M}$}}
\newcommand{\bbsN}{\mbox{$\mbox{\msbms N}$}} 
\newcommand{\bbsO}{\mbox{$\mbox{\msbms O}$}}
\newcommand{\bbsP}{\mbox{$\mbox{\msbms P}$}}
\newcommand{\bbsQ}{\mbox{$\mbox{\msbms Q}$}}
\newcommand{\bbsR}{\mbox{$\mbox{\msbms R}$}}
\newcommand{\bbsS}{\mbox{$\mbox{\msbms S}$}}
\newcommand{\bbsT}{\mbox{$\mbox{\msbms T}$}}
\newcommand{\bbsU}{\mbox{$\mbox{\msbms U}$}}
\newcommand{\bbsV}{\mbox{$\mbox{\msbms V}$}}
\newcommand{\bbsW}{\mbox{$\mbox{\msbms W}$}}
\newcommand{\bbsX}{\mbox{$\mbox{\msbms X}$}}
\newcommand{\bbsY}{\mbox{$\mbox{\msbms Y}$}}
\newcommand{\bbsZ}{\mbox{$\mbox{\msbms Z}$}}

\def\NNE{{\nb^\ast}}
\def\seqe#1{{{(#1_n)}_{n\in\NNE}}}
\def\TF{{{\rm T}\Phi}}
\def\rF{{\cal F}}
\def\rE{{\cal E}}
\def\rK{{\cal K}}
\def\rH{{\cal H}}
\def\ecarte{{\vphantom{\summ_i^n}}}
\newtheorem{remark}{Remark}
\newtheorem{theo}{Theorem}[section]
\newtheorem{pr}{Proposition}[section]
\newtheorem{cor}{Corollaire}[section]
\newtheorem{lem}{Lemme}[section]
\newtheorem{defn}{D\'{e}finition}[section]

\title{\Large\bf Statistical properties of Pauli matrices\\ going
through noisy channels} 
\author{Stéphane Attal and Nadine Guillotin-Plantard
\thanks{Universit\'e de Lyon, Universit\'e Lyon 1, Institut Camille
Jordan, U.M.R. C.N.R.S 5208, 43 bld du 11 novembre 1918, 69622  
Villeurbanne Cedex, e-mail: nadine.guillotin@univ-lyon1.fr;
attal@math.univ-lyon1.fr}} 
\maketitle
\begin{abstract}
We study the statistical properties of the triplet
$(\sigma_x,\sigma_y,\sigma_z)$ of Pauli matrices  going through a
sequence of noisy channels, modeled by the repetition of a general,
trace-preserving, completely positive map. We show a non-commutative
central limit theorem for the distribution of this triplet, which
shows up a 3-dimensional Brownian motion in the limit with a
non-trivial covariance matrix. We also prove a large deviation
principle associated to this convergence, with an explicit rate
function depending on the stationary state of the noisy channel.
\end{abstract}
\section{Introduction}
In quantum information theory one of the most important question is to
understand and to control the way a quantum bit is modified when
transmitted through a quantum channel. It is well-known that realistic
transmission channels are not perfect and that they distort the
quantum bit they transmit. This transformation of the quantum state is
represented by the action of a completely positive map. These are the
so-called noisy channels.

The purpose of this article is to study the action of the repetition
of a general  
completely positive map on basic observables. Physically, this model
can be thought of as the sequence of transformations of small identical
pieces of noisy channels on a qubit. It can also be thought of as a
discrete approximation of the more realistic model of a quantum bit
going through a semigroup of completely positive maps (a Lindblad
semigroup).  

As basic observables, we consider the triplet
$(\sigma_x,\sigma_y,\sigma_z)$ of Pauli matrices. Under the repeated
action of the completely positive map, they behave as a 3-dimensional
quantum random walk. The aim of this article is to study  the statistical
 properties of this
quantum random walk. 

Indeed, for any initial density matrix $\rho_{in}$, we study  the
statistical properties of the empirical average of the Pauli matrices
in the successive states $\Phi^n (\rho_{in}), n\geq 0$ where $\Phi$ is
some completely positive and trace-preserving map describing our
quantum channel. Quantum Bernoulli random walks studied by Biane in \cite{Bia} 
corresponds to the case where $\Phi$ is the identity map. Biane \cite{Bia} proved an invariance principle for this quantum random walk when $\rho_{in}=\frac{1}{2} I$.  

 This article is organized as follows. In section two
we describe the physical and mathematical setup. In section three we
establish a functional central limit theorem for the empirical average of the
quantum random walk associated to the Pauli matrices generalizing Biane's result \cite{Bia}. This central
limit theorem involves a 3-dimensional Brownian motion in the limit,
whose covariance matrix is non-trivial and depends explicitly on the
stationary state of the noisy channel. In section four,
we apply our central limit theorem to some explicit cases, in
particular to the King-Ruskai-Szarek-Werner representation of
completely positive and trace-preserving maps in $M_2(\bbC)$. This
allows us to compute the limit Brownian motion for the most well-know
quantum channels: the depolarizing channel, the phase-damping channel,
the amplitude-damping channel. Finally, in the last section, a large
deviation principle for the empirical average is proved. 

\section{Model and notations} 
Let $M_2(\bbC)$ be the set of $2\times 2$ matrices with complex
coefficients. The set of  $2\times 2$ self-adjoint matrices forms a
four dimensional real vector subspace of $M_2(\bbC)$. A convenient
basis ${\cal B}$ is given by the following matrices 

$$I =\left(\begin{array}{cc}
            $1$ & $0$\\
            $0$ & $1$
           \end{array}  \right) \ \ \sigma_x = \left(\begin{array}{cc}
            $0$ & $1$\\
            $1$ & $0$
           \end{array}  \right) \ \   \sigma_y= \left(\begin{array}{cc}
            $0$ & $-i$\\
            $i$ & $0$
           \end{array}  \right) \ \     \sigma_z=\left(\begin{array}{cc}
            $1$ & $0$\\
            $0$ & $-1$
           \end{array}  \right) $$   
where $\sigma_x, \sigma_y, \sigma_z$ are the traditional Pauli
matrices, they satisfy the commutation relations: 
$[\sigma_x, \sigma_y]=2i\sigma_z$, and those obtained by cyclic
permutations of $\sigma_x$, $\sigma_y$, $\sigma_z$. 
A state on $M_2(\bbC)$ is given by a density matrix (i.e. a positive
semi-definite matrix with trace one) which we will suppose to be of
the form  
$$\rho=\left(\begin{array}{ll}
\alpha & \ \beta \\
\bar{\beta}\  & 1-\alpha
\end{array}
\right)
$$
where $0\leq \alpha \leq 1$ and $|\beta|^2\leq \alpha(1-\alpha)$. The
noise coming from interactions between the qubit states and the
environment is represented by the action of a completely positive and
trace-preserving map $\Phi: M_2(\bbC)\rightarrow M_2(\bbC).$  

Let $M_1,\ M_2,\ldots,\ M_k,\ldots$ be infinitely many copies of
$M_2(\bbC)$. For each given state $\rho$, we consider the algebra 
$${\cal M}_{\rho}=M_1\otimes M_2\otimes\ldots\otimes M_k\otimes\ldots$$
where the product is taken in the sense of $W^{\ast}$-algebra with respect to the product state
$$
\omega=\rho\otimes\Phi(\rho)\otimes\Phi^2(\rho)\otimes\ldots
\otimes\Phi^k(\rho)\otimes\ldots.
$$  
Our main hypothesis is the following. 
We assume that for any state $\rho$, the sequance $\Phi^n(\rho)$
converges to a stationary state $\rho_{\infty}$, which we write as 
$$
\rho_{\infty}=\left(\begin{array}{ll}
\alpha_{\infty} & \ \beta_{\infty} \\
\overline{\beta}_{\infty}\  & 1-\alpha_{\infty}
\end{array}
\right)
$$
where $0\leq \alpha_{\infty} \leq 1$ and $|\beta_{\infty}|^2\leq
\alpha_{\infty}(1-\alpha_{\infty})$.\\*

Put 
$$
v_1=2\Re(\beta_{\infty}), v_2=-2\Im(\beta_{\infty}),
v_3=2\alpha_{\infty}-1.
$$
For every $k\ge 1$, we define
$$  x_k= I\otimes\ldots\otimes I\otimes(\sigma_{x}-v_1\ I)\otimes I\otimes\ldots$$
$$  y_k= I\otimes\ldots\otimes I\otimes(\sigma_{y}-v_2\ I)\otimes I\otimes\ldots$$
$$  z_k= I\otimes\ldots\otimes I\otimes(\sigma_{z}-v_3\ I)\otimes I\otimes\ldots$$
where each $(\sigma_{.}-v_{.}\ I)$ appears on the $k^{th}$ place.\\
For every $n\ge 1$, put
$$ X_n=\sum_{k=1}^n x_k,\  Y_n=\sum_{k=1}^n y_k,\  Z_n=\sum_{k=1}^n z_k$$
with initial conditions
$$X_0=Y_0=Z_0=0.$$ 
The integer part of a real $t$ is denoted by $[t]$. To each process we
associate a continuous time normalized process denoted by 
$$
X_{t}^{(n)}=n^{-1/2}X_{[nt]},\ Y_{t}^{(n)}=n^{-1/2}Y_{[nt]},\
Z_{t}^{(n)}=n^{-1/2}Z_{[nt]}.
$$

\section{A central limit theorem}

The aim of our article is to study the asymptotical properties of the
quantum process $(X^{(n)}_t, Y^{(n)}_t, Z^{(n)}_t)$ when $n$ goes to
infinity. This process being truly non-commutative, there is no hope
to obtain an asymptotic behaviour in the classical sense. 

For any polynomial $P=P(X_1,X_2,\ldots,X_m)$ of $m$ variables, we
denote by $\widehat{P}$ the {\it totally symmetrized polynomial} of
$P$ obtained by symmetrizing each monomial in the following way:  
$$
X_{i_1} X_{i_2}\ldots X_{i_k} \longrightarrow \frac{1}{k!}
\sum_{\sigma\in S_k} X_{i_{\sigma(1)}} \ldots X_{i_{\sigma(k)}}
$$ 
where $S_k$ is the group of permutations of $\{1,\ldots,k\}$.
 
\begin{theo}\label{tcl}
Assume that
\begin{equation*}
{\bf (A)}\ \ \ \ \ \ \ \ \ \ \  \Phi^{n}(\rho)= \rho_{\infty}+ o(\frac{1}{\sqrt{n}}).
\end{equation*}
Then, for any polynomial $P$ of $3m$  variables, for any $(t_1,\ldots,t_m)$
such that $0\leq t_1<t_2<\ldots<t_m$, the following convergence holds: 
$$
\lim_{n\rightarrow +\infty}
w\left[\widehat{P}(X_{t_1}^{(n)},Y_{t_1}^{(n)},Z_{t_1}^{(n)},\ldots,
X_{t_m}^{(n)},Y_{t_m}^{(n)},Z_{t_m}^{(n)})\right]
$$
$$
=\bbE\left[P(B_{t_1}^{(1)},B_{t_1}^{(2)},B_{t_1}^{(3)},\ldots
,B_{t_m}^{(1)},B_{t_m}^{(2)},B_{t_m}^{(3)})\right]
$$
where $(B_{t}^{(1)},B_{t}^{(2)},B_{t}^{(3)})_{t\ge 0}$ is a three-dimensional centered Brownian motion with covariance matrix $Ct$,
where
$$
C=\left(\begin{array}{lll}
1-v_{1}^2 & -v_1 v_2 &  -v_1 v_3 \\
-v_1 v_2 & 1-v_{2}^2 &  -v_2 v_3 \\
-v_1 v_3 &  -v_2 v_3 &  1-v_{3}^2
\end{array}
\right).
$$
\end{theo}
\noindent{\bf Remark~:}\\*
Theorem \ref{tcl} has to be compared with the quantum central limit theorem obtained 
in \cite{Gir} and \cite{Petz}. In our case, the state under which the convergence holds does not need to be an 
infinite tensor product of states. 
We also give here a functional version of the central limit theorem. Finally, in \cite{Gir} (see Remark 3 p.131), the limit is described as a so-called quasi-free state in quantum mechanics. We prove in Theorem \ref{tcl} that the limit is 
real Gaussian for the class of totally symmetrized polynomials.  

\noindent {\it Proof:}\\
Let $m\geq 1$ and $(t_0,t_1,\ldots,t_m)$ such that $t_0=0<t_1<t_2<\ldots<t_m$.
The polynomial
$P(X_{t_1}^{(n)},Y_{t_1}^{(n)},Z_{t_1}^{(n)},\ldots,X_{t_m}^{(n)},Y_{t_m}^{(n)},Z_{t_m}^{(n)})$
can be rewritten as a polynomial function $Q$ 
of the increments: $X_{t_1}^{(n)},\ Y_{t_1}^{(n)},\ Z_{t_1}^{(n)},\
X_{t_2}^{(n)}-X_{t_1}^{(n)},\ Y_{t_2}^{(n)}-Y_{t_1}^{(n)},\
Z_{t_2}^{(n)}-Z_{t_1}^{(n)},\ldots, 
\ X_{t_m}^{(n)}-X_{t_{m-1}}^{(n)},
\ Y_{t_{m}}^{(n)}-Y_{t_{m-1}}^{(n)}, \ Z_{t_{m}}^{(n)}-Z_{t_{m-1}}^{(n)}$.\\*
A monomial of $Q$ is a product of the form $A_{i_1}\ldots A_{i_k}$ for
some distinct $i_1,\ldots, i_k$ in $\{1,\ldots,m\}$ where $A_{i}$ is a
product depending only on  
the increments  $X_{t_i}^{(n)}-X_{t_{i-1}}^{(n)},\
Y_{t_i}^{(n)}-Y_{t_{i-1}}^{(n)},\
Z_{t_i}^{(n)}-Z_{t_{i-1}}^{(n)}$. Since the $A_i$'s are commuting
variables, the totally symmetrized polynomial of the monomial
$A_{i_1}\ldots A_{i_k}$ is equal to the product  $\widehat{A_{i_1}}
\ldots \widehat{A_{i_k}}$. Consequently, it is enough to prove the
theorem for any polynomial $A_i$. \\* 
Let $i\ge 1$ fixed, for every $\nu_1,\nu_2,\nu_3\in\bbR$, we begin by
determining the asymptotic distribution of the linear combination 
\begin{equation}\label{cl}
(\nu_1^2+\nu_2^2+\nu_3^2)^{-1/2}\Big(\nu_1
(X_{t_i}^{(n)}-X_{t_{i-1}}^{(n)})+ \nu_2
(Y_{t_i}^{(n)}-Y_{t_{i-1}}^{(n)}) +\nu_3
(Z_{t_i}^{(n)}-Z_{t_{i-1}}^{(n)})\Big) 
\end{equation}
 which can be rewritten as
$$\frac{1}{\sqrt{n}}\sum_{k=[nt_{i-1}]+1}^{[nt_{i}]}\left(\frac{\nu_1
x_k+ \nu_2 y_k+\nu_3 z_k}{\sqrt{\nu_1^2+\nu_2^2+\nu_3^2}} \right).$$ 
Consider the matrix
\begin{eqnarray*}
A &=& \frac{1}{\sqrt{\nu_1^2+\nu_2^2+\nu_3^2}} \Big(\nu_1
(\sigma_x-v_1 I) + \nu_2 (\sigma_y-v_2 I) +\nu_3 (\sigma_z-v_3
I)\Big)\\ 
&=&\frac{1}{\sqrt{\nu_1^2+\nu_2^2+\nu_3^2}}
\left(\begin{array}{ll}
-\nu_1 v_1-\nu_{2}v_2+\nu_3(1-v_3) & \ \ \nu_{1}-i\nu_{2} \\
\ \ \ \nu_{1}+i\ \nu_{2} & -\nu_1 v_1-\nu_{2}v_2-\nu_3(1+v_3)
\end{array}
\right)
\end{eqnarray*}
which we denote by
$$
 \left(\begin{array}{ll}
a_{1} & a_{3} \\
\overline{a}_{3} & a_{2} 
\end{array}
\right)\,, 
$$
$a_1,a_2\in\bbsR, a_3\in\bbsC.$

{}From assumption ${\bf (A)}$ we can write, for every $n\geq 0$
$$
\Phi^{n}(\rho)=\left(\begin{array}{ll}
\alpha_{\infty}+\phi_{n}(1) & \beta_{\infty}+\phi_{n}(2) \\
\overline{\beta}_{\infty}+\phi_{n}(3) & 1-\alpha_{\infty}+\phi_{n}(4)
\end{array}
\right) $$
where each sequence $(\phi_n(i))_{n}$ satisfies:
$\phi_n(i)=o(1/\sqrt{n})$.

Let $k\ge 1$, the expectation and the variance of $A$ in the state
$\Phi^{k}(\rho)$ are respectively equal to 
$$
\mbox{Trace}(A \Phi^{k}(\rho))
$$
and
$$ 
\mbox{Trace}(A^2 \Phi^{k}(\rho))-\mbox{Trace}(A \Phi^{k}(\rho))^2.
$$
If both following conditions are satisfied:
\begin{equation}\label{cond1}
\sum_{k=[nt_{i-1}]+1}^{[nt_i]}\mbox{Trace}(A \Phi^{k}(\rho))=o(\sqrt{n})
\end{equation}
and
\begin{equation}\label{cond2}
\lim_{n\ra +\infty}\,
\frac{1}{n}\sum_{k=[nt_{i-1}]+1}^{[nt_i]}[\mbox{Trace}(A^2
\Phi^{k}(\rho))-\mbox{Trace}(A \Phi^{k}(\rho))^2]=a(t_i-t_{i-1}), 
\end{equation}
then (see Theorem 2.8.42 in \cite{duflo}) the asymptotic distribution
of (\ref{cl}) is the Normal distribution ${\cal N}(0,a(t_i-t_{i-1})),
a>0$.

Let us first prove (\ref{cond1}).
For every $k\ge 1$, a simple computation gives 
$$
\mbox{Trace}(A \Phi^{k}(\rho))=[a_{1}\alpha_{\infty}+
a_{3}\bar{\beta}_{\infty}+ \overline{a}_{3} \beta_{\infty} +a_2
(1-\alpha_{\infty})+o(1/\sqrt{n})]=o(1/\sqrt{n}),
$$
hence
\begin{eqnarray*}
\sum_{k=[nt_{i-1}]+1}^{[nt_i]}\mbox{Trace}(A \Phi^{k}(\rho))&
=&\sum_{k=[nt_{i-1}]+1}^{[nt_i]}o(1/\sqrt{n})=o(\sqrt n). 
\end{eqnarray*}
This gives (\ref{cond1}).

Let us prove (\ref{cond2}).
Note that the sequence $(\mbox{Trace}(A \Phi^{n}(\rho)))_n$
converges to 0 , as $n$ tends to infinity. As a consequence, it is enough to prove
that 
$$
\frac{1}{n}\sum_{k=[nt_{i-1}]+1}^{[nt_i]}\mbox{Trace}(A^2 \Phi^{k}(\rho))$$
converges to a strictly positive constant.
A straightforward computation gives
\begin{eqnarray*}
& &\lim_{n\ra
+\infty}\frac{1}{n}\sum_{k=[nt_{i-1}]+1}^{[nt_i]}\mbox{Trace}(A^2
\Phi^{k}(\rho))\\ 
&=& a_1^2 \alpha_\infty+ a_2^2\
(1-\alpha_{\infty})+|a_3|^2+(a_1+a_2)(a_3\bar{\beta}_{\infty}
+\bar{a}_3{\beta}_{\infty}) \\ 
&=&\frac{(t_i-t_{i-1})}{\nu_1^2+\nu_2^2+\nu_3^2}\Big[\nu_1^2(1-v_{1}^2)
+\nu_2^2(1-v_{2}^2)+\nu_3^2(1-v_{3}^2) \\ 
&& - 2\nu_1\nu_2v_{1}v_{2}-2\nu_1\nu_3 v_{1}v_{3}-2\nu_2\nu_3 v_{2}v_{3}\Big].
\end{eqnarray*}
This means that, for every $\nu_1,\nu_2,\nu_3\in\bbR$, for any $p\geq 1$, the expectation 
\begin{eqnarray*}
w\Big[\Big(\nu_1 (X_{t_i}^{(n)}-X_{t_{i-1}}^{(n)})+ \nu_2 (Y_{t_i}^{(n)}-Y_{t_{i-1}}^{(n)}) +\nu_3 (Z_{t_i}^{(n)}-Z_{t_{i-1}}^{(n)})\Big)^{p}\Big]
\end{eqnarray*}
converges to 
$$\bbE\Big[\Big(\nu_1 (B^{(1)}_{t_i}-B^{(1)}_{t_{i-1}})+ \nu_2
(B_{t_i}^{(2)}-B_{t_{i-1}}^{(2)}) +\nu_3
(B_{t_i}^{(3)}-B_{t_{i-1}}^{(3)})\Big)^{p}\Big]\,,$$
where $(B^{(1)}_t,B^{(2)}_t,B^{(3)}_t)$ is a 3-dimensional Brownian
motion with the announced covariance matrix.

The polynomial
\begin{eqnarray*}
\Big(\nu_1 (X_{t_i}^{(n)}-X_{t_{i-1}}^{(n)})+ \nu_2
(Y_{t_i}^{(n)}-Y_{t_{i-1}}^{(n)}) +\nu_3
(Z_{t_i}^{(n)}-Z_{t_{i-1}}^{(n)})\Big)^{p} 
\end{eqnarray*}
can be developed as the sum
\begin{eqnarray*}
\sum_{0\leq p_1+p_2\leq p}  \nu_1^{p_1} \nu_2^{p_2}
\nu_3^{p-p_1-p_2}\sum_{\cal P} S_1 S_2\ldots S_{p} 
\end{eqnarray*}
where the summation in the last sum runs over all partitions ${\cal
P}=\{A,B,C\}$ of $\{1,\ldots,p\}$ such that  
$|A|=p_1,|B|=p_2,|C|=p-p_1-p_2$, with the convention:
$$S_{j}= \left\{ \begin{array}{lll}
X_{t_i}^{(n)}-X_{t_{i-1}}^{(n)} & \mbox{ if } & j \in A \\
Y_{t_i}^{(n)}-Y_{t_{i-1}}^{(n)} & \mbox{ if } & j \in B  \\
Z_{t_i}^{(n)}-Z_{t_{i-1}}^{(n)} & \mbox{ if } & j \in C\,.
\end{array}
\right.
$$
The expectation under $w$ of the above expression converges to the
corresponding expression involving the expectation ($\bbE[\,\cdot\,]$)
of the Brownian motion $(B^{(1)}_t,B^{(2)}_t,B^{(3)}_t)$. As this
holds for any $\nu_1,\nu_2,\nu_3\in\bbR$, we deduce that
 $w[\sum_{\cal P} S_1 S_2\ldots S_{p}]$ converges to the corresponding
 expectation for the Brownian motion.

We can conclude the proof by noticing that $\widehat{A_i}$ can be
written, modulo multiplication by a constant, as  $\sum_{\cal P} S_1
S_2\ldots S_{p}$ for some $p$. 
\epreuve

\medskip
Let us discuss the class of polynomials for which Theorem 3.1
holds. In the particular case when the map $\Phi$ is the identity map
and $\rho=1/2 I$ (in that case $v_i=0$ for $i=1,2,3$ and $C=I$), Biane
\cite{Bia} proved the convergence of the expectations in Theorem
\ref{tcl} for any polynomial in $3m$ non-commuting variables. It is
a natural question to ask whether our result holds for any polynomial
$P$ instead of $\widehat{P}$, or at least for a larger class. 

Let us give
an example of a polynomial for 
which the convergence in our setting does not hold.     
Take $P(X,Y)=XY$. From Theorem \ref{tcl}, the expectation under the state $\omega$ of
$$X_{t}^{(n)}Y_{t}^{(n)}+Y_{t}^{(n)}X_{t}^{(n)}$$
 converges as $n\rightarrow +\infty$ to $2\ \bbE[B_{t}^{(1)}B_{t}^{(2)}]$.\\
Since we have the following commutation relations 
\begin{equation*}
[(\sigma_x-v_1\ I),(\sigma_y-v_2\ I)]=2i \sigma_z,\ \ \ [(\sigma_y-v_2\ I),(\sigma_z-v_3\ I)]=2i \sigma_x
\end{equation*}
  and  
  $$[(\sigma_z-v_3\ I),(\sigma_x-v_1\ I)]=2i \sigma_y,$$
we deduce that
$$[X_{t}^{(n)},Y_{t}^{(n)}]=2i n^{-1/2}\ Z_{t}^{(n)} +2it v_3\ I,\ \
[Y_{t}^{(n)},Z_{t}^{(n)}]=2i n^{-1/2}\ X_{t}^{(n)} +2it v_1\ I$$ 
and
\begin{equation}\label{rc}
[Z_{t}^{(n)},X_{t}^{(n)}]=2i n^{-1/2}\ Y_{t}^{(n)} +2it v_2\ I.
\end{equation}
Then the expectation under the state $\omega$ of
$$X_{t}^{(n)}Y_{t}^{(n)} = \frac{1}{2} \Big[ \widehat{P}(X,Y)+[X_{t}^{(n)},Y_{t}^{(n)}]\Big]$$
 converges to 
 $\bbE[B_{t}^{(1)}B_{t}^{(2)}]+itv_3\neq\bbE[B_{t}^{(1)}B_{t}^{(2)}],$ 
if $v_3$ is non zero. \\

\medskip
Furthermore, by considering the polynomial $P(X,Y)=XY^3+Y^3X$, it is
possible to show that the convergence in Theorem \ref{tcl} can not be
enlarged to the class of symmetric polynomials.  Straightforward
computations gives that $P(X,Y)$ can be rewritten as 
$$\widehat{X Y^3}+\widehat{YX^3}+\frac{3}{4}[X,Y] (Y^2 -X^2) +\frac{1}{2}(Y[X,Y]Y -X[X,Y]X) 
+\frac{1}{4}(Y^2-X^2) [X,Y]$$
so the expectation $w[P(X_{t}^{(n)},Y_{t}^{(n)})]$ converges as $n$
tends to $+\infty$ to $$\bbE[P(B_t^{(1)},B_t^{(2)})]+3iv_3 t
(v_1^2-v_2^2)$$ 
which is not equal to $\bbE[P(B_t^{(1)},B_t^{(2)})]$ if $v_3\neq 0$
and $|v_1|\neq |v_2|$.

\smallskip
In the following corollary we give a condition under which the
convergence in Theorem \ref{tcl} holds for any polynomial in $3m$
non-commuting variables. 

\begin{cor}
In the case when $\rho_{\infty}$ is equal to $\frac{1}{2} I$, the
convergence holds for any polynomial $P$ in $3m$ non-commuting
variables, i.e. for every $t_1<t_2<\ldots<t_m$, the following
convergence holds: 
$$\lim_{n\rightarrow +\infty}
w\left[P(X_{t_1}^{(n)},Y_{t_1}^{(n)},Z_{t_1}^{(n)},
\ldots,X_{t_m}^{(n)},Y_{t_m}^{(n)},Z_{t_m}^{(n)})\right]$$  
$$=\bbE\left[P(B_{t_1}^{(1)},B_{t_1}^{(2)},B_{t_1}^{(3)},\ldots
,B_{t_m}^{(1)},B_{t_m}^{(2)},B_{t_m}^{(3)})\right]$$ 
where $(B_{t}^{(1)},B_{t}^{(2)},B_{t}^{(3)})_{t\ge 0}$ is a
three-dimensional centered Brownian motion with covariance matrix $t
I_3$. 
\end{cor}

\noindent {\it Proof:}\\
We consider the polynomials of the form 
$S = \frac{1}{N}\sum_{\cal P} S_1 S_2\ldots S_{p_1+p_2+p_3}$
where the summation is done over all partitions ${\cal P}=\{A,B,C\}$
of the set $\{1,\ldots,p_1+p_2+p_3\}$ such that  
$|A|=p_1,|B|=p_2,|C|=p_3$, with the convention:
$$S_{j}= \left\{ \begin{array}{lll}
X_{t_i}^{(n)}-X_{t_{i-1}}^{(n)} & \mbox{ if } & j \in A \\
Y_{t_i}^{(n)}-Y_{t_{i-1}}^{(n)} & \mbox{ if } & j \in B  \\
Z_{t_i}^{(n)}-Z_{t_{i-1}}^{(n)} & \mbox{ if } & j \in C
\end{array}
\right.
$$ and $N$ is the number of terms in the sum.

{}From Theorem \ref{tcl} the expectation under the state $w$ of $S$ converges to 
$$\bbE\Big[ \displaystyle\prod_{j=1}^{3} (B^{(j)}_{t_i} - B^{(j)}_{t_{i-1}})^{p_{j}}\Big].$$
Using the commutation relations (\ref{rc}) with the $v_i$'s being  all
equal to zero,
monomials of $S$ differ of each other by $n^{-1/2}$ times a polynomial
of total degree less or equal to $(p_1+p_2+p_3)-1$. It is easy to
conclude by  induction.   
\epreuve

\section{Examples}
\subsection{King-Ruskai-Szarek-Werner's representation}
The set of  $2\times 2$ self-adjoint matrices forms a four dimensional
real vector subspace of $M_2(\bbC)$. A convenient basis of this space
is given by ${\cal B}=\{I,\sigma_x,\sigma_y,\sigma_z\}$. Each state
$\rho$ on $M_2(\bbC)$ can then be written as
$$\rho=\frac{1}{2}\left(\begin{array}{ll}
1+z & x-i\ y \\
x+i\ y & 1-z
\end{array}
\right)
$$
where $x,y,z$ are reals such that $x^2+y^2+z^2\le 1$. Equivalently, in the basis ${\cal B}$, 
$$\rho=\frac{1}{2}(I+x\ \sigma_x+y\ \sigma_y+z\ \sigma_z)$$ with
$x,y,z$ defined above. Thus, the set of density matrices can be
identified with the unit ball in $\bbR^3$. 
The pure states, that is, the
ones for which $x^2+y^2+z^2=1$, constitutes the Bloch sphere. 

The noise
coming from interactions between the qubit states and the environment
is represented by the action of a completely positive and
trace-preserving map $\Phi: M_2(\bbC)\rightarrow M_2(\bbC).$ Kraus and
Choi \cite{Choi, kra1, kra2} gave an abstract representation of these
particular maps in terms of Kraus operators: There exists at most four
matrices $L_i$ such that for any density matrix $\rho$, 
$$\Phi(\rho)= \sum_{1\leq i\leq 4} L_i^{*} \rho L_i$$
with $\sum_i L_i L_i^{*} = I$.
The matrices $L_i$ are usually called the {\it Kraus operators} of
$\Phi$. This representation is unique up to a unitary
transformation. Recently, King, Ruskai et al \cite{Rus1, Rus2}
obtained a precise characterization of completely positive and
trace-preserving maps from $M_2(\bbC)$ as follows. The map $\Phi:
M_2(\bbC)\rightarrow M_2(\bbC)$ being linear and preserving the trace,
it can be represented as a unique $4\times4$-matrix in the basis   
$\cal B$ given by
$$\left(\begin{array}{ll}
1 & {\bf 0} \\
{\bf t} & {\bf T}
\end{array}
\right)
$$
with ${\bf 0}=(0,0,0)$, ${\bf t}\in\bbR^3$ and $\bf T$ a real
$3\times3$-matrix. King, Ruskai et al \cite{Rus1, Rus2} proved that
via changes of basis, this matrix can be reduced  to  
\begin{equation}\label{matrix}
T=\left(\begin{array}{llll}
1 & 0 & 0 & 0 \\
t_1 & \lambda_1 & 0 & 0\\
t_2 & 0 & \lambda_2 & 0\\
t_3 & 0 & 0 & \lambda_3\\
\end{array}
\right)
\end{equation}
Necessary and sufficient conditions under which the map $\Phi$ with
reduced matrix $T$ for which  
$|t_3|+|\lambda_3|\leq 1$ is completely positive are (see \cite{Rus2})  
\begin{equation}\label{Cond1}
(\lambda_1+\lambda_2)^2\leq
(1+\lambda_3)^2-t_3^2-(t_1^2+t_2^2)\left(\frac{1+\lambda_3\pm
t_3}{1-\lambda_3\pm t_3}\right)\leq (1+\lambda_3)^2-t_3^2   
\end{equation} 
\begin{equation}\label{Cond2}
(\lambda_1-\lambda_2)^2\leq
(1-\lambda_3)^2-t_3^2-(t_1^2+t_2^2)\left(\frac{1-\lambda_3\pm
t_3}{1+\lambda_3\pm t_3}\right)\leq (1-\lambda_3)^2-t_3^2   
\end{equation} 
\begin{eqnarray}\label{Cond3}
& &\left[1-(\lambda_1^2+\lambda_2^2+\lambda_3^2)-(t_1^2+t_2^2+t_3^2)\right]^2\nonumber\\
& & \ \ \ \geq 4
\left[\lambda_1^2(t_1^2+\lambda_2^2)+\lambda_2^2(t_2^2+\lambda_3^2)+
\lambda_3^2(t_3^2+\lambda_1^2)-2\lambda_1\lambda_2\lambda_3\right]\,.   
\end{eqnarray} 
We now apply Theorem \ref{tcl} in this setting. Let $\Phi$ be a
completely positive and trace preserving map with matrix $T$ given in
(\ref{matrix}), with coefficients $t_i, \lambda_i, i=1,2,3$ satisfying
conditions (\ref{Cond1}), (\ref{Cond2}) and (\ref{Cond3}). Moreover,
we assume that $|\lambda_i|<1, i=1,2,3$. 
For every $n\geq 0$,
$$\Phi^{n}(\rho)=\frac{1}{2}\left(\begin{array}{ll}
1+\phi_{n}(3) & \phi_{n}(1)-i\ \phi_{n}(2) \\
\phi_{n}(1)+i\ \phi_{n}(2) & 1-\phi_{n}(3)
\end{array}
\right) $$
where  the sequences $(\phi_n(i))_{n\ge 0}$, $i=1,2,3$ satisfy the induction relations:
$$ 
\phi_n(i)=\lambda_i \phi_{n-1}(i)+t_i.
$$
with initial conditions $\phi_0(1)=x, \phi_0(2)=y$ and $\phi_0(3)=z$.
Explicit formulae can easily be obtained. We get,
for every $n\geq 0$,
$$\phi_n(1)=\Big(x-\frac{t_1}{1-\lambda_1}\Big) \lambda_1^{n}+\frac{t_1}{1-\lambda_1}$$
$$\phi_n(2)=\Big(y-\frac{t_2}{1-\lambda_2}\Big) \lambda_2^{n}+\frac{t_2}{1-\lambda_2}$$
$$\phi_n(3)=\Big(z-\frac{t_3}{1-\lambda_3}\Big) \lambda_3^{n}+\frac{t_3}{1-\lambda_3}.$$
Hence, for any state $\rho$, for any $n\geq 1$, 
$$\Phi_n(\rho)= \rho_{\infty}+o(|\lambda|_{max}^n)$$ 
where $|\lambda|_{max}=\displaystyle\max_{i=1,2,3}|\lambda_i|$ and 
$$\rho_{\infty}=\left(\begin{array}{ll}
\alpha_{\infty} & \ \beta_{\infty} \\
\overline{\beta}_{\infty}\  & 1-\alpha_{\infty}
\end{array}
\right)
$$ 
with $\alpha_{\infty}=\frac{1}{2}\displaystyle\left(1+\frac{t_3}{1-\lambda_3}\right)$
and $\beta_{\infty}= \frac{1}{2}\displaystyle\left(\frac{t_1}{1-\lambda_1}-i\ \frac{t_2}{1-\lambda_2}\right)$.  Theorem \ref{tcl} applies with 
$\displaystyle v_i=\frac{t_i}{1-\lambda_i}, i=1,2,3$.

\smallskip
We now give some examples of well-known quantum channels.  For each of
them we give their Kraus operators, their corresponding matrix $T$ in
the King-Ruskai-Szarek-Werner's representation, as well as the vector
$v=(v_1,v_2,v_3)$ and the covariance matrix $C$ obtained in Theorem
\ref{tcl}. 
It is worth noticing that if $\Phi$ is a {\it unital} map, i.e. such
that $\Phi(I) = I$, then the covariance matrix $C$ is equal to the
identity matrix $I_3$.  

\begin{enumerate}
\item{\it The depolarizing channel:}

Kraus operators: for some $0\leq p\leq 1$,
$$L_1=\sqrt{1-p}I, L_2=\sqrt{\frac{p}{3}} \sigma_x, L_3=\sqrt{\frac{p}{3}} \sigma_y,
L_4=\sqrt{\frac{p}{3}} \sigma_z.$$

King-Ruskai-Szarek-Werner's representation:
$$
T=\left(\begin{array}{llll}
\ \ \ 1 & \ \ 0 & \ \ 0 & \ \ \ 0 \\
\ \ \ 0 & 1-\frac{4p}{3} & \ \ 0 & \ \ \ 0 \\
\ \ \ 0 & \ \ 0 &  1-\frac{4p}{3} & \ \ \ 0 \\
\ \ \ 0 & \ \ 0 & \ \ 0 & 1-\frac{4p}{3} \\
\end{array}
\right)
$$

The vector $v$ is the null vector  and the covariance matrix $C$ in
this case is given by the identity matrix $I_3$.

\medskip
\item{\it Phase-damping channel:}

Kraus operators: for some $0\leq p\leq 1$,
$$L_1=\sqrt{1-p}\ I,
\  L_2=\sqrt{p} \left(\begin{array}{ll}
 1 & \ \ 0 \\
 0 & \ \ 0
\end{array}
\right), 
L_3=\sqrt{p} \left(\begin{array}{ll}
 0 & \ \ 0 \\
 0 & \ \ 1
\end{array}
\right)
$$

King-Ruskai-Szarek-Werner's representation:
$$
T=\left(\begin{array}{llll}
\ \ \ 1 & \ \ 0 & \ \ 0 & \ \ \ 0 \\
\ \ \ 0 & 1-p & \ \ 0 & \ \ \ 0 \\
\ \ \ 0 & \ \ 0 &  1-p & \ \ \ 0 \\
\ \ \ 0 & \ \ 0 & \ \ 0 &\ \ \  1  \\
\end{array}
\right)
$$

The vector $v$ is the null  vector  and the covariance matrix $C$ in
this case is given by $I_3$.

\medskip
\item{\it Amplitude-damping channel:} 

Kraus operators: for some $0\leq p\leq 1$,
$$L_1=\left(\begin{array}{ll}
 1 &  0 \\
 0 &  \sqrt{1-p}
\end{array}
\right),
L_2=\left(\begin{array}{ll}
 0 & \sqrt{p} \\
 0 &  0
\end{array}
\right)
$$

King-Ruskai-Szarek-Werner's representation:
$$
T=\left(\begin{array}{llll}
\ \ \ 1 & \ \ 0 & \ \ 0 & \ \ \ 0 \\
\ \ \ 0 &  \sqrt{1-p} & \ \ 0 & \ \ \ 0 \\
\ \ \ 0 & \ \ 0 &  \sqrt{1-p} & \ \ \ 0 \\
\ \ \ t & \ \ 0 & \ \ 0 & 1-p\\
\end{array}
\right)
$$

The vector $v$ is equal to $(0,0,1)$.
The covariance matrix in this case is given by 
$$C=\left(\begin{array}{lll}
1 & 0 & 0 \\
0 & 1 & 0 \\
0 & 0 & 0
\end{array}
\right)$$

\medskip
\item {\it Trigonometric parameterization: } 

Consider the particular Kraus operators
$$L_1=\Big[\cos(\frac{v}{2})\cos(\frac{u}{2})\Big]I+
\Big[\sin(\frac{v}{2})\sin(\frac{u}{2})\Big]\sigma_z$$ and
$$L_2=\Big[\sin(\frac{v}{2})\cos(\frac{u}{2})\Big]\sigma_x-i
\Big[\cos(\frac{v}{2})\sin(\frac{u}{2})\Big]\sigma_y.$$

King-Ruskai-Szarek-Werner's representation:
$$
T=\left(\begin{array}{llll}
\ \ \ 1 & \ \ 0 & \ \ 0 & \ \ \ 0 \\
\ \ \ 0 & \cos u & \ \ 0 & \ \ \ 0 \\
\ \ \ 0 & \ \ 0 & \cos v & \ \ \ 0 \\
\sin u\sin v & \ \ 0 & \ \ 0 & \cos u\cos v \\
\end{array}
\right)
$$

The vector $v$ is equal to $\displaystyle (0,0,\frac{\sin u\sin v}{1-\cos u\cos v}).$
The covariance matrix in this case is given by
$$C=\left(\begin{array}{lll}
1 & 0 &  \ \ 0 \\
0 & 1 &  \ \ 0 \\
0 & 0 &  1-v_3^2
\end{array}
\right)$$
with $\displaystyle v_3=\frac{\sin u\sin v}{1-\cos u\cos v}$.
\end{enumerate}

\subsection{CP map associated to a Markov chain}

With every Markov chain with two states and transition matrix given by
$$P=\left(\begin{array}{ll}
p & \ 1-p\\
q & \ 1-q
\end{array}
\right),\ p,q\in (0,1)
$$
is associated a completely positive and trace preserving map, denoted by $\Phi$, with the Kraus operators:
$$L_1=\left(\begin{array}{ll}
\sqrt{p} & \  \sqrt{1-p} \\
0 & \ \  \ \ 0
\end{array}
\right)=
\frac{\sqrt{p}}{2}(I+\sigma_z)+\frac{\sqrt{1-p}}{2}(\sigma_x+i\sigma_y)$$
 and
$$ L_2=\left(\begin{array}{ll}
\ 0 & \ \ \ \ 0 \\
\sqrt{q} & \sqrt{1-q}
\end{array}
\right)=\frac{\sqrt{1-q}}{2}(I-\sigma_z)+\frac{\sqrt{q}}{2}(\sigma_x-i\sigma_y).
$$
Let $\rho$ be the density matrix 
$$\frac{1}{2}\left(\begin{array}{ll}
1+z & x-i\ y \\
x+i\ y & 1-z
\end{array}
\right)
$$
where $x,y,z$ are reals such that $x^2+y^2+z^2\le 1$. The map $\Phi$
transforms the density matrix $\rho$ into  
a new one given by 
$$\Phi(\rho)= L_1^{*} \rho L_1+ L_2^{*} \rho L_2.$$
By induction, for every $n\geq 0$,
$$\Phi^{n}(\rho)=\left(\begin{array}{ll}
p_{n} & \ \ \ r_{n} \\
r_{n} & 1-p_{n}
\end{array}
\right) $$ where  the sequences $(p_n)_{n\ge 0}$, and $(r_n)_{n\ge
0}$ satisfy the recurrence relations: for every $n\geq 1$,
$$p_n=p_{n-1}(p-q)+q$$
and
$$r_n=\sqrt{q(1-q)} + p_{n-1}( \sqrt{p(1-p)} -\sqrt{q(1-q)})$$
 with the initial condition $p_0=(1+z)/2$. Assumption ${\rm (A)}$ is
 then clearly satisfied with 
 $$\rho_{\infty}=\frac{1}{1+q-p}\left(\begin{array}{ll}
q & \beta  \\
\beta\  & 1-p 
\end{array}
\right)$$
where $\beta= \Big[q\sqrt{p(1-p)} + (1-p)\sqrt{q(1-q)} \Big]$.
Then, applying Theorem \ref{tcl}, if $P$ is a polynomial of $3m$
non-commuting variables, for every $0<t_1<t_2<\ldots<t_m$, the 
following convergence holds
$$\lim_{n\rightarrow +\infty}
w\left[\widehat{P}(X_{t_1}^{(n)},Y_{t_1}^{(n)},Z_{t_1}^{(n)},
\ldots,X_{t_m}^{(n)},Y_{t_m}^{(n)},Z_{t_m}^{(n)})\right]$$ 
$$=\bbE\left[P(B_{t_1}^{(1)},B_{t_1}^{(2)},B_{t_1}^{(3)},
\ldots,B_{t_m}^{(1)},B_{t_m}^{(2)},B_{t_m}^{(3)})\right]$$ 
where $(B_{t}^{(1)},B_{t}^{(2)},B_{t}^{(3)})_{t\ge 0}$ is a
three-dimensional centered Brownian motion with Covariance matrix
$Ct$ where
$$C=\left(\begin{array}{lll}
1-v_{1}^2 &  0  &  -v_1 v_2 \\
\ \ \ 0  &  1  &  \ \ \ 0 \\
-v_1 v_2 &  0 &  1-v_{2}^2
\end{array}
\right)$$ 
with $$v_{1}=\frac{2}{1+q-p}[q\sqrt{p(1-p)} + (1-p)\sqrt{q(1-q)} ]$$
and $$v_{2}=\frac{p+q-1}{1+q-p}.$$

\section{Large deviation principle}
Let $\Gamma$ be a Polish space endowed with the Borel
$\sigma$-field ${\cal B}(\Gamma)$. A good {\it rate function} is a
lower semi-continuous function $\Lambda^{*}: \Gamma \ra
[0,\infty]$ with compact level sets $\{x; \Lambda^{*}(x)\leq
\alpha\}, \alpha\in [0,\infty[.$ Let $v=(v_{n})_{n}\uparrow\infty$
be an increasing sequence of positive reals. A sequence of random
variables $(Y_{n})_{n}$ with values in $\Gamma$ defined
 on a probability space $(\Omega, {\cF}, \bbP)$ is said to satisfy
 {\it a Large Deviation Principle} (LDP)
 with speed $v=(v_{n})_{n}$ and the good rate function $\Lambda^{*}$ if
for every Borel set $B\in{\cal B}(\Gamma)$,
\begin{eqnarray*}
-\inf_{x\in B^{o}}\Lambda^{*}(x)&\leq&\liminf_{n}\frac{1}{v_{n}}\log \bbP(Y_{n}\in B)\\
&\leq&\limsup_{n}\frac{1}{v_{n}}\log \bbP(Y_{n}\in B)\leq -\inf_{x\in\bar{B}}\Lambda^{*}(x).
\end{eqnarray*}
For every $k\ge 1$, we define
$$  \bar{x}_k= I\otimes\ldots\otimes I\otimes\sigma_{x}\otimes I\otimes\ldots$$
$$  \bar{y}_k= I\otimes\ldots\otimes I\otimes\sigma_{y}\otimes I\otimes\ldots$$
$$  \bar{z}_k= I\otimes\ldots\otimes I\otimes\sigma_{z}\otimes I\otimes\ldots$$
where each $\sigma_{.}$ appears on the $k^{th}$ place.\\
For every $n\ge 1$, we consider the processes
$$ \bar{X}_n=\sum_{k=1}^n \bar{x}_k,\  \bar{Y}_n=\sum_{k=1}^n
\bar{y}_k,\  \bar{Z}_n=\sum_{k=1}^n \bar{z}_k$$ 
with initial conditions
$$
\bar{X}_0=\bar{Y}_0=\bar{Z}_0=0.
$$
To each vector $\nu=(\nu_1, \nu_2 , \nu_3)\in\bbR^3$, we associate the
euclidean norm $||\nu ||=\sqrt{\nu_1^2+\nu_2^2+\nu_3^2}$ and $\bra
.,.\ket$ the corresponding inner product. 
\begin{theo}\label{th1}
Let $\Phi$ be a completely positive and trace-preserving map for which
there exists a state $$\rho_{\infty}=\left(\begin{array}{ll} 
\alpha_{\infty} & \ \beta_{\infty} \\
\overline{\beta}_{\infty}\  & 1-\alpha_{\infty}
\end{array}
\right)
$$ such that for any given state $\rho$,
$$\Phi^{n}(\rho)= \rho_{\infty}+o(1).
$$
For every $\nu=(\nu_1, \nu_2 , \nu_3)\in\bbR^{3,*}$, the sequence 
$$\Big(\displaystyle \frac{\nu_1 \bar{X}_n+\nu_2 \bar{Y}_n+\nu_3 \bar{Z}_n}{n}\Big)_{n\geq 1}
$$ 
satisfies a LDP with
speed $n$ and the good rate  function 
$$
I(x)=\left\{
\begin{array}{lll}
\frac{1}{2}\left[
\left(1+\frac{x}{\vert\vert\nu\vert\vert}\right)
\log\left( {{\vert\vert\nu\vert\vert+x}\over{\vert\vert\nu\vert\vert+ \bra
\nu,v\ket}}\right)\right.&\\
\ \ \ +  
\left.\left(1-\frac{x}{\vert\vert\nu\vert\vert}\right)
\log\left(\frac{\vert\vert\nu\vert\vert- x}{\vert\vert\nu\vert\vert- \bra
\nu,v\ket}\right)\right] & \mbox{ if }\ 
|x|<\vert\vert\nu\vert\vert.\\*
\\*
+\infty\ & \mbox{otherwise.} 
\end{array} \right.
$$
where $v_1=2\Re(\beta_{\infty}), v_2=-2\Im(\beta_{\infty}), v_3=2\alpha_{\infty}-1.$
\end{theo}
\noindent {\it Proof:}\\
The matrix
\begin{eqnarray*}
B:=\nu_1 \sigma_x+\nu_2 \sigma_y+\nu_3 \sigma_z &=&
\left(\begin{array}{ll}
\ \ \nu_3 & \ \ \nu_{1}-i\nu_{2} \\
\nu_{1}+i\ \nu_{2} & -\nu_3
\end{array}
\right)
\end{eqnarray*}
has two distinct eigenvalues $\pm ||\nu||$.\\*
For every $n\geq 0$, we can write 
$$\Phi^{n}(\rho)=\left(\begin{array}{ll}
\alpha_{\infty}+\phi_{n}(1) & \beta_{\infty}+\phi_{n}(2) \\
\overline{\beta}_{\infty}+\phi_{n}(3) & 1-\alpha_{\infty}+\phi_{n}(4)
\end{array}
\right) $$
where the four sequences $(\phi_n(i))_{n\ge 0}$ satisfy  $\phi_n(i)=o(1)$.

For any $k\ge 1$, the expectation of $B$ in the state $\Phi^{k}(\rho)$ is equal to
$$\mbox{Trace}(B\ \Phi^{k}(\rho))=\displaystyle \bra \nu, v\ket+\varepsilon_k,$$
with $\varepsilon_n=o(1)$. As a consequence, the distribution of $B$ is 
$$p_k(\vert\vert\nu\vert\vert)=\frac{1}{2} \left
[ 1+\frac{1}{\vert\nu\vert } (\bra \nu, v\ket+\varepsilon_k) \right]=1-p_k(-||\nu||).$$ 
Using the fact that $\nu_1 \bar{X}_n+\nu_2 \bar{Y}_n+\nu_3 \bar{Z}_n$
is the sum of $n$ commuting matrices, we get that  
\begin{eqnarray*}
& &\frac{1}{n} \log w\ (\exp{t (\nu_1 X_n+\nu_2 Y_n+\nu_3 Z_n)})\\
&  & \ \ \ = \frac{1}{n}  \sum_{k=1}^n \log\Big(e^{||\nu|| t} p_k(||\nu||) 
+e^{-||\nu|| t} (1-p_k(||\nu||))\Big)
\end{eqnarray*}
Since $\varepsilon_n=o(1)$, we obtain that
\begin{eqnarray*}
& &\lim_{n\rightarrow +\infty} \frac{1}{n} \log w\ (\exp{t (\nu_1 X_n+\nu_2 Y_n+\nu_3 Z_n)})\\
& & \ \ \ =\log\Big(\cosh{(||\nu|| t)}+\frac{\bra\nu, v\ket}{||\nu||} \sinh{(||\nu|| t)}\Big)\\
& &\ \ \ = \log\Big(\cosh{(||\nu|| t)}\Big) +\log \Big(1+
\frac{\bra\nu, v\ket}{||\nu||} \tanh{(||\nu|| t)} \Big).
\end{eqnarray*}
We denote by $\Lambda(t)$ this function of $t$.

For every $t \in \bbR$, the function $\Lambda$ is finite and
differentiable on $\bbR$, then, by G\"{a}rtner-Ellis Theorem (see
\cite{Dem}), the LDP holds with the good 
rate function 
$$I(x)=\sup_{t\in\bbsR} \{tx - \Lambda(t)\}.$$
A simple computation leads to the rate function given in the theorem.\\
\epreuve

\end{document}